\documentclass[sigconf]{acmart}

\AtBeginDocument{%
  }

\setcopyright{none}
\acmConference[Preprint]{}{2025}{}
\acmISBN{}
\acmDOI{}

\usepackage{booktabs}
\usepackage{needspace}
\usepackage{placeins}
\usepackage{tabularx}
\usepackage{siunitx}
\usepackage{graphicx}
\usepackage{float}
\usepackage{caption}
\usepackage[type={CC},modifier={by},version={4.0}]{doclicense}
\usepackage{array} 
\newcolumntype{L}{>{\large\raggedright\arraybackslash}p{6cm}}
\newcolumntype{C}{>{\centering\arraybackslash}X}          
\begin{document}

\title{Evaluating MCC for Low-Frequency Cyberattack Detection in Imbalanced Intrusion Detection Data}

\author{Prameshwar Thiyagarajan}
\affiliation{%
  \institution{Farmington High School}
  \city{Farmington}
  \state{Connecticut}
  \country{USA}
}
  
\author{Chad A. Williams}
\affiliation{%
  \institution{Central Connecticut State University}
  \department{Department of Computer Science}
  \city{New Britain}
  \state{Connecticut}
  \country{USA}
}

\renewcommand{\shortauthors}{Thiyagarajan}

\begin{abstract}
In many real-world network environments, several types of cyberattacks occur at very low rates compared to benign traffic, making them difficult for intrusion detection systems (IDS) to detect reliably. This imbalance causes traditional evaluation metrics, such as accuracy, to often overstate model performance in these conditions, masking failures on minority attack classes that are most important in practice. In this paper, we evaluate a set of base and meta classifiers on low-traffic attacks in the CSE-CIC-IDS2017 dataset and compare their reliability in terms of accuracy and Matthews Correlation Coefficient (MCC). The results show that accuracy consistently inflates performance, while MCC provides a more accurate assessment of a classifier's performance across both majority and minority classes. Meta-classification methods, such as LogitBoost and AdaBoost, demonstrate more effective minority class detection when measured by MCC, revealing trends that accuracy fails to capture. These findings establish the need for imbalance-aware evaluation and make MCC a more trustworthy metric for IDS research involving low-traffic cyberattacks.
\end{abstract}

\maketitle

\section{Introduction}

Intrusion detection systems (IDSs) are widely used in real-world networks to detect attacks \citep{diana2025overview}. However, real-world network traffic is highly imbalanced, with attacks accounting for only a fraction of benign traffic, making minority-attack detection challenging \citep{9103235}. 

\indent This challenge is made worse by how IDS performance is typically evaluated. IDS models are commonly assessed using accuracy, but this often exaggerates the performance when classes are imbalanced. As a result, a classifier might seem adequate while performing poorly on rare attacks, such as web-based or infiltration attacks. Even though several studies have explored how to improve a model's architecture or feature representations, not many have examined how evaluation metrics can influence conclusions about a model's effectiveness.

In this paper, we evaluate a set of base and meta-classifiers on low-traffic attacks in the CSE-CIC-IDS2017 dataset \citep{sharafaldin2018toward}. We compare performance using accuracy and the Matthews Correlation Coefficient (MCC), a metric designed to handle imbalanced data. Our results show that accuracy consistently overestimates classifier performance. In contrast, MCC provides a clearer view of strengths and weaknesses across both majority and minority attack classes, highlighting the need for IDS research to evaluate models using metrics that account for class imbalance rather than relying solely on accuracy.

\section{Related Work}

Machine learning has long been used to support cybersecurity tasks such as intrusion detection, anomaly detection, and malware analysis. Surveys of AI applications in cybersecurity highlight challenges, including high false-positive rates, evolving attack behaviors, and significant class imbalance in real-world network traffic \citep{74d266f9-c3e6-3ad1-b9d9-3959a92b6438}. Their paper underscores that the effectiveness of IDS models is not solely determined by algorithmic sophistication but by how well they handle imbalanced attack distributions, showing the need to examine model performance specifically on low-frequency attacks.

In response to this challenge, researchers have proposed a variety of modeling approaches to improve IDS performance. Deep hybrid architectures, such as CNN-LSTM and autoencoder combinations, are better at capturing both spatial and temporal structure in system-call sequences, underscoring the need for evaluation methods that treat all classes fairly \citep{b099020e-8b1a-3b32-b4f5-09705372afc7}. Other papers have focused more directly on imbalance mitigation using oversampling, embedded feature extraction, and stacked representations to improve the classification of minority attacks, but these studies show that traditional metrics like accuracy often obscure weaknesses in rare-attack detection \citep{talukder2024mlintrusion}. These findings suggest that architectural improvements alone cannot ensure reliable performance on minority classes.

Beyond IDS-specific approaches, the broader literature on imbalanced datasets highlights similar limitations. Surveys of imbalance methods note that standard classifiers tend to favor the majority class, thereby reducing their reliability against rare attacks \citep{ganganwar2012imbalanced}. Research on evaluation metrics in imbalanced settings shows that accuracy and ROC-based measures often misrepresent model effectiveness, and identifies MCC as a more reliable alternative\citep{Hancock_Khoshgoftaar_Johnson_2023}.Although F1-score is commonly used to balance precision and recall, prior work has shown that it can be misleading under extreme class imbalance because it ignores true negatives. In contrast, Matthews Correlation Coefficient (MCC) incorporates all four confusion matrix components and remains reliable even when the minority class is severely underrepresented\citep{chicco_jurman_2020_mcc}. This makes MCC more appropriate for evaluating low-frequency cyberattacks. Studies in intrusion detection have similar conclusions, noting that low-frequency attacks are often misclassified unless other balancing strategies or more appropriate evaluation metrics are used \citep{93def600-04c0-3fd2-96fa-848f68509e85}. These findings highlight the importance of metrics that remain stable and informative even under severe class imbalance.

Although prior research has explored advanced architectures, feature engineering, and imbalance-mitigation strategies, relatively few studies analyze how evaluation metrics themselves can shape conclusions about classifier performance. This research gap motivates the present work, which compares advanced meta-classifiers and baseline models across three low-frequency attacks—Heartbleed, Web Attacks, and Infiltration—in the CSE-CIC-IDS2017 dataset. By contrasting accuracy with MCC, the study demonstrates how metric choice can alter model rankings and provide a more reliable assessment of IDS performance on rare cyberattacks.

\section{Dataset and Preprocessing}

This paper uses three sampled subsets of the CICIDS2017 dataset corresponding to Heartbleed, Web Attack, and Infiltration to evaluate classifier performance. The CICIDS2017 dataset was developed to capture realistic network attack scenarios based on the most common attacks identified in the 2016 McAfee report, such as Web-based intrusions, Brute Force attempts, DoS/DDoS events, Botnet activity, Port Scanning, Infiltration, and Heartbleed. Network flows were collected for five consecutive days, with regular traffic recorded on Monday and attack scenarios executed on the remaining days. Each flow contains over 80 statistical and time-based features extracted using the CIC FlowMeter tool.

For this paper, the original dataset files were converted to ARFF format and downsampled using WEKA's SpreadSubsample filter, with a maximum class count of 50,000 instances, while retaining representative samples. No features were removed, and the data was not normalized or scaled. Each subset's attack instances were labeled as malicious, while all the remaining traffic was labeled as benign. For the Web Attack subset, the original attack subtypes (XSS, SQL Injection, and Brute Force Web) were merged into a single malicious label, while the Infiltration and Heartbleed subsets consisted of their respective attack types mapped directly to the Malicious class.

After labeling and sampling, the resulting subsets showed severe class imbalance. This imbalance reflects the distribution of the original CICIDS2017 traffic capture, where malicious attacks occur far less frequently than benign traffic, rather than being an artifact introduced by the sampling methodology. The Heartbleed subset contains 11 Heartbleed instances and 50,000 benign instances, making Heartbleed only 0.022\% of the dataset. The Infiltration subset contains 36 infiltration instances and 20,000 benign instances, with the attack class comprising 0.18\% of the data. The Web Attack subset contains 2,180 attack instances and 20,000 benign instances, with malicious samples accounting for 9.8\% of the dataset.These distributions show that instances of Heartbleed and Infiltration attacks are extremely underrepresented in the dataset, while Web Attack represents a minority class. This imbalance reflects the small volume of malicious traffic relative to benign traffic captured during data collection. This severe imbalance supports the use of metrics such as MCC, which are reliable even when accuracy can be misleading.

\section{Methodology}

\subsection{Classifiers evaluated}

To compare model performance under severe imbalance, this paper evaluates two categories of machine learning models: baseline classifiers and meta-classifiers. Baseline classifiers represent standard single-model approaches commonly used in intrusion detection research, such as Logistic Regression, Random Forests, Naïve Bayes, J48 (Decision Tree), and JRip. These models will serve as a reference point to evaluate how traditional classifiers handle low-frequency attacks.

Unlike baseline classifiers, meta-classifiers operate on top of other classifiers by adding extra steps such as resampling, feature selection, or combining multiple models. Instead of learning directly from the raw data, they modify or coordinate base classifiers to improve overall predictive performance. Many of the meta-classifiers used in this study are ensemble-based methods, meaning they combine multiple base learners through resampling, feature-subspace variation, or weighted aggregation to improve robustness under class imbalance. In this paper, the ensemble-based meta-classifiers are AdaBoostM1, LogitBoost, Bagging, RandomSubSpace, and RandomCommittee. These algorithms capture a range of ensemble and meta-level strategies to examine whether such approaches offer improved minority-class detection compared to baseline models.

This classifier group enables direct comparison between traditional single classifier approaches and meta-level approaches to address the research question of whether meta-classifiers achieve stronger MCC performance on rare cyberattacks.

In addition to comparing baseline and meta-level methods, this study also groups classifiers into six algorithmic families based on their underlying learning model:\textbf{Decision Trees, Rule-Based Learners, Lazy (Instance-Based) Methods, Bayesian Models, Meta/Ensemble Methods, and Function-Based Models}. Grouping classifiers like this allows performance trends to be observed across model types rather than individual algorithms, which is important under extreme class imbalance. This categorization also ensures that comparisons remain consistent across all three evaluated datasets (Infiltration, Web Attack, and Heartbleed), enabling to evaluate between classifiers in the results tables that follow.

\subsection{Experimental Setup}

All experiments in the study were conducted using the WEKA software toolkit. Each classifier was evaluated using 10-fold cross-validation, a standard method for estimating performance in supervised learning. Cross-validation partitions the data into 10 sets, trains on 9 of them, tests on the remaining set, and repeats the process 10 times. This procedure ensures that each instance is used exactly once for testing, reducing the variance introduced by the extreme imbalance across all three subsets.

\subsection{Evaluation Metrics}

Although WEKA reports multiple metrics such as precision, recall, error rate, and ROC area, this study will focus on accuracy and MCC. Accuracy is widely reported in IDS evaluations, but it is unreliable under severe imbalance, as the majority class dominates the dataset. MCC, on the other hand, incorporates all four confusion matrices (TP, TN, FP, FN) and continues to reflect actual model performance even when the minority class is only a fraction of the dataset. For clarity, the Matthews Correlation Coefficent (MCC) is defined as: 

\begin{equation}
MCC = \frac{TP \times TN - FP \times FN}{\sqrt{(TP + FP)(TP + FN)(TN + FP)(TN + FN)}}
\end{equation} 

This formula highlights the sensitivity of MCC to all four confusion metrics, which makes it more reliable than metrics such as accuracy or F1-score under extreme class imbalance. A model may score a very high accuracy yet have an MCC score of 0 which indicates that the model predicted every instance as benign and was unable to detect a singular attack showing how accuracy can be misleading under extreme class imbalance.

Given that the Heartbleed, Web Attacks, and Infiltration datasets have attack percentages ranging from 0.022\% to 9.8\%, relying solely on Accuracy can create a false sense of model effectiveness. MCC, therefore, serves as the primary metric for determining minority-attack detection performance, while Accuracy is reported alongside MCC to illustrate how the metrics diverge under imbalance.

\section{Results}

\subsection{Infiltration Results}

The infiltration dataset contains 36 attack instances and 20,000 benign instances, with the widest difference between accuracy and MCC among the attacks. Almost every classifier achieved an accuracy of 99.82\% to 99.999\% suggesting that the classifiers are performing very well. In reality, these high accuracies are misleading and do not reflect the whole picture. Multiple classifiers failed to identify a single infiltration attempt, resulting in zero true positives and an MCC score of 0, despite high accuracy. These classifiers appeared effective when judged solely by accuracy, yet were unable to detect any attacks.

However, other models demonstrated strong minority class performance which accuracy was unable to highlight. For example, DecisionTable correctly identified all 36 infiltration instances and achieved an MCC score of 0.986 (98.6\%), even though its accuracy (99.995\%) was slightly lower than that of weaker models. These differences help show that accuracy is unreliable under extreme imbalance, as it cannot distinguish between models that detect infiltration attacks and those that miss all of them.  MCC, on the other hand, clearly separates effective classifiers from ineffective classifiers and provides a more accurate representation of model performance.

\subsection{Web Attack Results}

The Web Attack dataset had a larger minority class, with 2,180 attack instances and 20,000 benign instances, resulting in less imbalance than in the infiltration and Heartbleed datasets. As a result, many classifiers achieved both high accuracy and high MCC, with numerous baseline and meta-classifiers exceeding 99\% accuracy while also producing MCC values in the 0.97–0.99 range. The higher attack percentage aligns accuracy more closely with minority class performance.

However, despite these high overall accuracies, closer inspection using MCC showed substantial differences in how well models identified Web attacks. Some classifiers with accuracy around 99.8\% showed noticeably lower MCC values, suggesting that misclassifying Web attacks persisted despite strong overall performance. Ensemble-based meta-classifiers, such as Bagging, RandomSubSpace, and RandomCommittee, tended to produce the highest MCC values and demonstrated more consistent detection of Web attacks compared to weaker baseline models. These results show that while accuracy becomes more informative at moderate imbalance levels, MCC remains a more reliable metric for assessing the detection of minority cyberattacks.

\subsection{Heartbleed Results}

The Heartbleed dataset had the most significant imbalance among all datasets, with only 11 attack instances and 50,000 benign instances. Almost all the classifiers achieved an accuracy of around 99.97\% with several models reporting perfect accuracy under cross-validation. However, the apparent uniformity suggested by accuracy covered substantial variation in actual detection capability. Some classifiers achieved MCC scores near 1.0, indicating near-perfect Heartbleed detection, while others produced much lower MCC values despite reporting accuracy similar to the strongest models.

This occurs because a classifier may be unable to identify Heartbleed attacks while still maintaining almost perfect accuracy, given the volume of benign traffic. As in the previous datasets, MCC provided a more reliable assessment of classifier performance on rare attacks, whereas accuracy failed to differentiate between models that succeeded and those that struggled.

\subsection{Summary}

Across all three attack types, the results show a consistent pattern. Accuracy remains high across nearly all classifiers, regardless of how well the model detects minority attacks. Infiltration and Heartbleed, in particular, demonstrate that the benign class dominates accuracy and therefore cannot show meaningful variation in how well classifiers identify attacks. However, MCC consistently distinguishes between strong and weak classifiers across all attack types, revealing the actual performance that accuracy masks. Meta-classifiers generally achieved higher MCC values than baseline models, suggesting that ensemble-based approaches are more effective at identifying low-frequency attacks. The overall results confirm that MCC provides the most dependable measure of classifier effectiveness when evaluating intrusion detection models on heavily imbalanced network traffic.

\subsection{Cross-Attack Aggregate Ranking}
\needspace{10\baselineskip}
While Sections 5.1–5.3 presented classifier performance separately for Infiltration, Web Attacks,
and Heartbleed, evaluating each dataset independently does not reveal which models perform
best overall. To measure this, the Mean MCC was computed for every classifier by averaging its
MCC scores across all three datasets. Table~\ref{tab:top5-mean-mcc} summarizes the top five
classifiers based on mean MCC.

\FloatBarrier 

\begin{table}[H]
\centering
\captionsetup{font=normalsize, labelfont=bf} 
\caption{\textbf{Top 5 Classifiers by Mean MCC Across Three Attack Types}
(Classifier families shown in parentheses)}
\label{tab:top5-mean-mcc}

\scriptsize
\setlength{\tabcolsep}{2.5pt}
\renewcommand{\arraystretch}{1.05}

\resizebox{\columnwidth}{!}{%
\begin{tabular}{lccc>{\bfseries}c}
\toprule
\textbf{Classifier} & \textbf{Inf.} & \textbf{Web} & \textbf{Heart.} & \textbf{Mean} \\
\midrule
FilteredClassifier (Meta/Ens.) & 87.3 & 99.3 & 100.0 & 95.53 \\
ClassificationViaRegression (Meta/Ens.) & 94.4 & 99.5 & 90.9 & 94.93 \\
JRip (Rule-Based) & 87.0 & 99.1 & 95.3 & 93.80 \\
RandomForest (Decision Tree) & 85.0 & 99.4 & 95.3 & 93.23 \\
RandomSubSpace (Meta/Ens.) & 83.4 & 99.4 & 95.7 & 92.83 \\
\bottomrule
\end{tabular}%
}
\end{table}

\noindent
The ranking shows that ensemble/meta-based classifiers dominate overall performance, with 
FilteredClassifier and ClassificationViaRegression achieving the highest aggregated MCC, at 
95.53\% and 94.93\% respectively. JRip and RandomForest also demonstrate strong overall performance, ranking among the top classifiers by average MCC even though they are baseline models. While en-sembles tend to dominate under extreme imbalance, these results suggest that robust baseline models are still capable of detecting low-frequency attacks effectively, making them viable alternatives.
\section{Discussion}

The results across all three datasets show that accuracy is not a reliable indicator of classifier performance under severe imbalance. Infiltration and Heartbleed show this most clearly, as they had a greater imbalance than Web Attack. Multiple classifiers achieved an accuracy of around 99.8\% yet were unable to identify a single attack. These findings highlight that the majority class dominates accuracy and provides limited information about detection capability in highly skewed settings.

Unlike accuracy, MCC provided a more accurate assessment of classifier performance across all attack types. Since MCC incorporates all components of the confusion matrix, it captures differences that accuracy does not. For example, models such as DecisionTable and ClassificationViaRegression achieved high MCC values on the Infiltration dataset by correctly identifying most or all attack instances, whereas others with similar accuracy produced MCC values near zero. The same pattern appeared in Web Attacks and Heartbleed, indicating that MCC consistently distinguishes between classifiers that learn minority-class patterns and those that do not.

The results also show that classifier families behave differently under imbalance. Meta-classifiers and ensemble-based models tended to achieve higher MCC values than baseline classifiers, although some baseline models performed competitively on specific datasets. This suggests that ensemble methods can help mitigate imbalance, but strong performance is not limited to any single model type. Instead, the evaluation metric plays a central role in determining whether these differences are visible. This trend is further reflected in the overall ranking results, where models such as FilteredClassifier, ClassificationViaRegression, JRip, RandomForest, and RandomSubSpace ranked as the top performers across all three attacks. While ensemble-style methods appear advantageous overall, the presence of strong baseline models among the top results reinforces that algorithm choice alone is not sufficient; the evaluation metric ultimately determines which models are truly effective under imbalance.

The Web Attacks dataset, which had a higher proportion of attack instances, showed closer alignment between accuracy and MCC. This indicates that the divergence between the two metrics is influenced by the severity of imbalance rather than the classifiers themselves. As the minority class becomes smaller, accuracy becomes less reliable, while MCC maintains sensitivity and gives informative information.

Overall, the findings demonstrate the importance of using evaluation metrics that remain informative under skewed class distributions. For intrusion detection tasks that involve rare attacks, imbalance-aware measures like MCC offer a clearer view of model performance and provide a more reliable basis for comparison than accuracy alone.

\section{Limitations}

One limitation is that the study relies on three sampled subsets of CICIDS2017 rather than the full dataset. While this choice helps run experiments on standard hardware, sampling may reduce the diversity of benign behavior and attack traffic, limiting the generalizability of the findings. Also, the subsets are from an older dataset, and more recent datasets like CICIDS2021 or UNSW-NB15, to determine whether the observed differences between MCC and Accuracy persist in larger, more varied traffic environments. 

The second limitation is the study's binary labeling. All attack subtypes were merged into a single malicious class, which does not accurately reflect the full complexity of real-world intrusion detection, where systems must distinguish among multiple attack families. With different courses, analysis of MCC, accuracy, and other evaluation metrics may yield different results. 

The third limitation is that the paper only evaluates baseline and meta classifiers in WEKA. While WEKA covers a wide range of machine learning approaches, it does not include modern deep learning architectures such as CNN–LSTM hybrids or autoencoder-based anomaly detection systems. Extending the MCC-based analysis to deep learning models would help determine whether the divergence between MCC and Accuracy persists across more complex models. 

Finally, this work uses 10-fold cross-validation for evaluation. Even though 10-fold cross-validation is the standard, it assumes balanced class representation across folds, which may not hold in extreme class imbalance. Other evaluation strategies, such as repeated cross-validation, stratified bootstrapping, or an alternative validation scheme, to assess whether MCC remains stable across different sampling strategies.

Together, these limitations highlight the need for broader experimentation across larger datasets, more complex models, and alternative evaluation schemes. Addressing these areas in future work can help determine the extent to which the findings generalize to modern intrusion-detection and strengthen the understanding of MCC’s role in evaluating rare cyberattacks.

\section{Conclusion}

This study evaluated baseline and meta-classifiers on three low-frequency attacks (Infiltration, Web Attacks, and Heartbleed) using sampled subsets of the CICIDS2017 dataset. The results show that, although nearly all classifiers achieved high accuracy across all three attacks, none detected a single malicious instance. Matthews Correlation Coefficient (MCC), however, reliably distinguishes between models that have successfully identified minority attacks and those that did not, offering a more reliable and informative measure in imbalanced conditions.

Across all three attack types, ensemble-based meta-classifiers generally achieved higher MCC scores and were more consistent at identifying minority attacks than baseline models. This suggests that ensemble-based approaches are better suited for detecting rare cyber threats. These results reinforce the limitations of accuracy as an evaluation metric in intrusion detection and highlight the value of MCC as a more dependable alternative when attack frequencies are low.

In conclusion, this work demonstrates that the choice of metric can meaningfully affect conclusions about classifier performance in IDS research. By showing how MCC reveals performance differences that accuracy masks, the study highlights the importance of using imbalance-aware evaluation metrics when assessing models on real-world, skewed network traffic. Future IDS research can benefit from adopting MCC alongside accuracy to obtain a more complete and reliable understanding of classifier effectiveness against rare cyberattacks.

The findings of this study have direct relevance for real-world intrusion detection deployment, where rare attacks often pose the highest risk yet remain difficult to identify. MCC-based evaluation could be integrated into SOC model selection pipelines to prevent the deployment of classifiers that appear effective under accuracy but fail on low traffic attacks. Future work may expand this analysis using larger unsampled datasets, modern deep learning IDS architectures, or real-time traffic streams to determine whether the advantages of MCC and ensemble methods persist at scale. Integrating MCC-driven evaluation into operational IDS monitoring systems may ultimately support earlier detection of low-frequency cyberattacks that traditional metrics overlook.

\bibliographystyle{ACM-Reference-Format}
\bibliography{custom} 

@techreport{74d266f9-c3e6-3ad1-b9d9-3959a92b6438,
 URL = {http://www.jstor.org/stable/resrep22692},
 author = {Loaiza, Francisco L. and Birdwell, John D. and Kennedy, George L. and  Visser, Dale},
 institution = {Institute for Defense Analyses},
 title = {Utility of Artificial Intelligence and Machine Learning in Cybersecurity},
 urldate = {2025-11-24},
 year = {2019}
}

@book{b099020e-8b1a-3b32-b4f5-09705372afc7,
 URL = {https://jstor.org/stable/community.39573397},
 author = {Osamor, Frances Nkiru and Wellman, Briana},
 publisher = {University of the District of Columbia},
 title = {Advancing anomaly detection in cybersecurity : deep learning-based hybrid models for system call sequence analysis},
 address = {Washington, D.C.},
 year = {2024},
 urldate = {2025-11-24}
}

@article{talukder2024mlintrusion,
  author       = {Talukder, Md. Alamin and Islam, Md. Manowarul and Uddin, Md Ashraf and Hasan, Khondokar Fida and Sharmin, Selina and Alyami, Salem A. and Moni, Mohammad Ali},
  title        = {Machine learning-based network intrusion detection for big and imbalanced data using oversampling, stacking feature embedding and feature extraction},
  journal      = {Journal of Big Data},
  volume       = {11},
  number       = {33},
  year         = {2024},
  doi          = {10.1186/s40537-024-00886-w},
  numpages     = {44},
  url          = {https://doi.org/10.1186/s40537-024-00886-w},
}

@article{ganganwar2012imbalanced,
  author    = {Ganganwar, Vaishali},
  title     = {An Overview of Classification Algorithms for Imbalanced Datasets},
  journal   = {International Journal of Emerging Technology and Advanced Engineering},
  volume    = {2},
  number    = {4},
  pages     = {42--47},
  year      = {2012},
  month     = apr,
  issn      = {2250-2459},
  url       = {https://www.researchgate.net/profile/Vaishali-Ganganwar/publication/292018027_An_overview_of_classification_algorithms_for_imbalanced_datasets/links/58c7707a458515478dc4c68b/An-overview-of-classification-algorithms-for-imbalanced-datasets.pdf},
}

@article{Hancock_Khoshgoftaar_Johnson_2023, title={Evaluating classifier performance with highly imbalanced Big Data}, volume={10}, DOI={10.1186/s40537-023-00724-5}, abstractNote={Using the wrong metrics to gauge classification of highly imbalanced Big Data may hide important information in experimental results. However, we find that analysis of metrics for performance evaluation and what they can hide or reveal is rarely covered in related works. Therefore, we address that gap by analyzing multiple popular performance metrics on three Big Data classification tasks. To the best of our knowledge, we are the first to utilize three new Medicare insurance claims datasets which became publicly available in 2021. These datasets are all highly imbalanced. Furthermore, the datasets are comprised of completely different data. We evaluate the performance of five ensemble learners in the Machine Learning task of Medicare fraud detection. Random Undersampling (RUS) is applied to induce five class ratios. The classifiers are evaluated with both the Area Under the Receiver Operating Characteristic Curve (AUC), and Area Under the Precision Recall Curve (AUPRC) metrics. We show that AUPRC provides a better insight into classification performance. Our findings reveal that the AUC metric hides the performance impact of RUS. However, classification results in terms of AUPRC show RUS has a detrimental effect. We show that, for highly imbalanced Big Data, the AUC metric fails to capture information about precision scores and false positive counts that the AUPRC metric reveals. Our contribution is to show AUPRC is a more effective metric for evaluating the performance of classifiers when working with highly imbalanced Big Data.}, number={1}, journal={Journal of Big Data}, author={Hancock, John T. and Khoshgoftaar, Taghi M. and Johnson, Justin M.}, year={2023}, pages={42} }

@phdthesis{93def600-04c0-3fd2-96fa-848f68509e85,
 URL = {https://jstor.org/stable/community.38760643},
 author = {Abedzadeh, Najmeh},
 school = {The Catholic University of America},
 advisor = {Jacobs, Matthew},
 publisher = {The Catholic University of America},
 title = {Implementing a New Algorithm to Balance and Classify the Imbalanced Intrusion Detection System Datasets},
 year = {2022},
 urldate = {2025-11-24}
}

@inproceedings{sharafaldin2018toward,
  author    = {Sharafaldin, Iman and Lashkari, Arash H. and Ghorbani, Ali A.},
  title     = {Toward Generating a New Intrusion Detection Dataset and Intrusion Traffic Characterization},
  booktitle = {Proceedings of the 4th International Conference on Information Systems Security and Privacy (ICISSP)},
  pages     = {108--116},
  year      = {2018},
  doi       = {10.5220/0006639801080116},
  publisher={SciTePress Set{\'u}bal},
  address = {Funchal, Madeira, Portugal},

  url       = {https://www.scitepress.org/papers/2018/66398/66398.pdf}
}

@article{diana2025overview,
  author  = {Diana, L. and Dini, Pierpaolo and Paolini, Davide },
  title   = {Overview on Intrusion Detection Systems for Computers and Network Security},
  journal = {Computers},
  VOLUME = {14},
  NUMBER = {3},
  ARTICLE-NUMBER = {87},
  NUMPAGES = {44},
  year    = {2025},
  url     = {https://www.mdpi.com/2073-431X/14/3/87}
}

@ARTICLE{9103235,
  author={Xiong, Bo and Mahoney, Eric and Lo, Joe F. and Fang, Qiyin},
  journal={IEEE Journal of Selected Topics in Quantum Electronics}, 
  title={A Frequency-Domain Optofluidic Dissolved Oxygen Sensor With Total Internal Reflection Design For in Situ Monitoring}, 
  year={2021},
  volume={27},
  number={4},
  pages={1-7},
  doi={10.1109/JSTQE.2020.2997810}}

@article{chicco_jurman_2020_mcc,
  author       = {Chicco, Davide and Jurman, Giuseppe},
  title        = {The advantages of the {Matthews} correlation coefficient ({MCC}) over {F1} score and accuracy in binary classification evaluation},
  journal      = {BMC Genomics},
  volume       = {21},
  pages        = {6},
  year         = {2020},
  doi          = {10.1186/s12864-019-6413-7},
  url          = {https://doi.org/10.1186/s12864-019-6413-7}
}

\newpage

\begin{table*}[t]
\centering
\Large
\setlength{\tabcolsep}{9pt}
\renewcommand{\arraystretch}{1.3}

\caption{\Large \textbf{Classifier Performance on the Infiltration Subset of CICIDS2017}}
\label{tab:infiltration-results}

\begin{tabular}{L c c c}
\toprule
\textbf{\Large Classifier} & \textbf{\Large Accuracy (\%)} & \textbf{\Large Recall (\% Attack)} & \textbf{\Large MCC} \\ 
\midrule

\multicolumn{4}{l}{\textbf{\textit{\Large Decision Trees}}}\\
J48 & 99.9152 & 61.1 & 73.3\\
RandomForest & 99.9501 & 72.2 & 85.0\\
RandomTree & 99.9052 & 63.9 & 71.1\\
REPTree & 99.9451 & 72.2 & 83.4\\
DecisionStump & 99.8203 & 0.0 & 0.0\\

\midrule
\multicolumn{4}{l}{\textbf{\textit{\Large Rule-Based Learners}}}\\
JRip & 99.9551 & 83.3 & 87.0\\
OneR & 99.9750 & 100.0 & 93.7\\
PART & 99.9351 & 69.44 & 80.2\\
DecisionTable & 99.9950 & 100.0 & 98.6\\
ZeroR & 99.8203 & 0.0 & 0.0\\

\midrule
\multicolumn{4}{l}{\textbf{\textit{\Large Lazy / Instance-Based}}}\\
IBk & 99.9251 & 63.9 & 76.6\\

\midrule
\multicolumn{4}{l}{\textbf{\textit{\Large Bayesian Models}}}\\
BayesNet & 99.9138 & 80.56 & 22.7\\
NaiveBayesMultinomialText & 99.8203 & 0.0 & 0.0\\

\midrule
\multicolumn{4}{l}{\textbf{\textit{\Large Meta/Ensemble Methods}}}\\
Bagging & 99.9301 & 63.9 & 78.2\\
AdaBoostM1 & 99.9301 & 61.1 & 78.1\\
ClassificationViaRegression & 99.9301 & 94.4 & 94.4\\
AttributeSelectedClassifier & 99.8852 & 61.1 & 65.8\\
FilteredClassifier & 99.9551 & 86.1 & 87.3\\
IterativeClassifierOptimizer & 99.9401 & 66.7 & 81.6\\
LogitBoost & 99.9401 & 66.7 & 81.6\\
RandomCommittee & 99.9401 & 72.2 & 81.9\\
MultiScheme & 99.8203 & 0.0 & 0.0\\
CVParameterSelection & 99.8203 & 0.0 & 0.0\\
Stacking & 99.8203 & 0.0 & 0.0\\
Vote & 99.8203 & 0.0 & 0.0\\
WeightedInstancesHandlerWrapper & 99.8203 & 0.0 & 0.0\\
InputMappedClassifier & 99.8203 & 0.0 & 0.0\\
RandomSubSpace & 99.9451 & 72.2 & 83.4\\

\midrule
\multicolumn{4}{l}{\textbf{\textit{\Large Function-Based}}}\\
SGDText & 99.8203 & 0.0 & 0.0\\
VotedPerceptron & 99.8203 & 0.0 & 0.0\\

\bottomrule
\end{tabular}
\end{table*}

\newpage

\begin{table*}[t]
\centering
\Large
\setlength{\tabcolsep}{9pt}
\renewcommand{\arraystretch}{1.3}

\caption{\Large \textbf{Classifier Performance on the Web Attack Subset of CICIDS2017}}
\label{tab:web-results}

\begin{tabular}{L c c c}
\toprule
\textbf{\Large Classifier} & \textbf{\Large Accuracy (\%)} & \textbf{\Large Recall (\% Attack)} & \textbf{\Large MCC} \\
\midrule

\multicolumn{4}{l}{\textbf{\textit{\Large Decision Trees}}}\\
J48 & 99.8422 & 99.2 & 99.1\\
RandomForest & 99.8918 & 99.1 & 99.4\\
RandomTree & 99.7565 & 98.8 & 98.6\\
REPTree & 99.9053 & 99.6 & 99.5\\
DecisionStump & 94.6979 & 83.4 & 73.0\\

\midrule
\multicolumn{4}{l}{\textbf{\textit{\Large Rule-Based Learners}}}\\
JRip & 99.8377 & 99.2 & 99.1\\
OneR & 99.3251 & 98.1 & 96.1\\
PART & 99.9008 & 99.5 & 99.4\\
DecisionTable & 98.2642 & 97.9 & 91.0\\
ZeroR & 99.8203 & 0.0 & 0.0\\

\midrule
\multicolumn{4}{l}{\textbf{\textit{\Large Lazy / Instance-Based}}}\\
IBk & 99.8370 & 99.3 & 99.1\\

\midrule
\multicolumn{4}{l}{\textbf{\textit{\Large Bayesian Models}}}\\
BayesNet & 97.8796 & 97.5 & 89.2\\
NaiveBayesMultinomialText & 99.8203 & 0.0 & 0.0\\

\midrule
\multicolumn{4}{l}{\textbf{\textit{\Large Meta/Ensemble Methods}}}\\
Bagging & 99.9594 & 99.7 & 99.6\\
AdaBoostM1 & 97.3150 & 83.4 & 84.2\\
ClassificationViaRegression & 99.9098 & 99.6 & 99.5\\
AttributeSelectedClassifier & 99.7701 & 99.1 & 98.7\\
FilteredClassifier & 99.8780 & 99.3 & 99.3\\
IterativeClassifierOptimizer & 98.8320 & 89.6 & 93.1\\
LogitBoost & 98.8320 & 89.6 & 93.1\\
RandomCommittee & 99.8740 & 99.1 & 99.3\\
MultiScheme & 99.8203 & 0.0 & 0.0\\
CVParameterSelection & 99.8203 & 0.0 & 0.0\\
Stacking & 99.8203 & 0.0 & 0.0\\
Vote & 99.8203 & 0.0 & 0.0\\
WeightedInstancesHandlerWrapper & 99.8203 & 0.0 & 0.0\\
InputMappedClassifier & 99.8203 & 0.0 & 0.0\\
RandomSubSpace & 99.9010 & 99.1 & 99.4\\

\midrule
\multicolumn{4}{l}{\textbf{\textit{\Large Function-Based}}}\\
SGDText & 99.8203 & 0.0 & 0.0\\
VotedPerceptron & 96.5239 & 83.7 & 80.6\\

\bottomrule
\end{tabular}
\end{table*}

\newpage

\begin{table*}[t]
\centering
\Large
\setlength{\tabcolsep}{9pt}
\renewcommand{\arraystretch}{1.3}

\caption{\Large \textbf{Classifier Performance on the Heartbleed Subset of CICIDS2017}}
\label{tab:heartbleed-results}

\begin{tabular}{L c c c}
\toprule
\textbf{\Large Classifier} & \textbf{\Large Accuracy (\%)} & \textbf{\Large Recall (\% Attack)} & \textbf{\Large MCC} \\ 
\midrule

\multicolumn{4}{l}{\textbf{\textit{\Large Decision Trees}}}\\
J48 & 99.994 & 90.9 & 87.0\\
RandomForest & 99.998 & 90.9 & 95.3\\
RandomTree & 100.0 & 100.0 & 100.0\\
REPTree & 99.996 & 100.0 & 92.0\\
DecisionStump & 100.0 & 100.0 & 100.0\\

\midrule
\multicolumn{4}{l}{\textbf{\textit{\Large Rule-Based Learners}}}\\
JRip & 99.998 & 90.9 & 95.3\\
OneR & 99.992 & 90.9 & 83.6\\
PART & 99.996 & 90.9 & 87.0\\
DecisionTable & 99.992 & 90.9 & 83.6\\
ZeroR & 99.978 & 0.0 & 0.0\\

\midrule
\multicolumn{4}{l}{\textbf{\textit{\Large Lazy / Instance-Based}}}\\
IBk & 99.998 & 90.9 & 95.3\\

\midrule
\multicolumn{4}{l}{\textbf{\textit{\Large Bayesian Models}}}\\
BayesNet & 99.976 & 100.0 & 69.1\\
NaiveBayesMultinomialText & 99.978 & 0.0 & 0.0\\

\midrule
\multicolumn{4}{l}{\textbf{\textit{\Large Meta/Ensemble Methods}}}\\
Bagging & 100.0 & 100.0 & 100.0\\
AdaBoostM1 & 100.0 & 100.0 & 100.0\\
ClassificationViaRegression & 99.996 & 90.9 & 90.9\\
AttributeSelectedClassifier & 99.996 & 90.9 & 90.9\\
FilteredClassifier & 100.0 & 100.0 & 100.0\\
IterativeClassifierOptimizer & 100.0 & 100.0 & 100.0\\
LogitBoost & 100.0 & 100.0 & 100.0\\
RandomCommittee & 99.998 & 90.9 & 95.3\\
MultiScheme & 99.978 & 0.0 & 0.0\\
CVParameterSelection & 99.978 & 0.0 & 0.0\\
Stacking & 99.978 & 0.0 & 0.0\\
Vote & 99.978 & 0.0 & 0.0\\
WeightedInstancesHandlerWrapper & 99.978 & 0.0 & 0.0\\
InputMappedClassifier & 99.978 & 0.0 & 0.0\\
RandomSubSpace & 99.998 & 100.0 & 95.7\\

\midrule
\multicolumn{4}{l}{\textbf{\textit{\Large Function-Based}}}\\
SGDText & 99.978 & 0.0 & 0.0\\
VotedPerceptron & 99.978 & 0.0 & 0.0\\

\bottomrule
\end{tabular}
\end{table*}

\end{document}